\documentclass[a4paper,11pt]{article}
\usepackage{jheppub}
\usepackage[dvipsnames]{xcolor}
\usepackage{multirow}
\usepackage{physics}
\usepackage{tabulary}
\usepackage{listings}
\usepackage{lineno}

\usepackage{graphicx,color}

\usepackage{slashed}
\usepackage{comment}

\allowdisplaybreaks[4]

\newcommand{\akh}[1]{{\color{blue} #1}}

\newcommand{\be}{\begin{equation}}
\newcommand{\ee}{\end{equation}}
\newcommand{\ba}{\begin{eqnarray}}
\newcommand{\ea}{\end{eqnarray}}
\newcommand{\mq}{m_q}

\def\KITA{Institute for Theoretical Particle Physics, KIT, Wolfgang-Gaede-Straße 1, 76131, Karlsruhe, Germany}

\def\Siegen{Theoretische Physik 1, Center for Particle Physics Siegen, Universit\"at Siegen, 57068 Siegen, Germany}

\def\SantaBarbara{Kavli Institute for Theoretical 
Phyiscs, University of California, Santa Barbara}
\def\FTPI{Fine Theoretical Physics Institute and School of Physics and Astronomy, University of Minnesota}

\preprint{
\begin{flushright}
SI-HEP-2025-26,
TTP25-051,
P3H-25-104
\end{flushright}
}

\title{Non-perturbative effects in  Higgs boson decays to electroweak vector bosons and photons }
\author[a]{Alexander Khodjamirian,}
\author[b]{Kirill Melnikov,}
\author[c]{Arkady Vainshtein}

\emailAdd{kirill.melnikov@kit.edu,
khodjamirian@physik.uni-siegen.de,
vainshte@umn.edu}

\affiliation[a]{\Siegen}
\affiliation[b]{\KITA}
\affiliation[c]{\SantaBarbara; \FTPI}

\abstract{ 
We estimate the magnitude of the leading \emph{non-perturbative}  QCD 
 corrections  to the decays of  the Higgs boson to the  $\gamma Z$ and $\gamma \gamma$ final states.
 These  corrections  originate from the 
  light-quark contributions  to such  decays. 
 We show that  the non-perturbative effects  
are suppressed by  the small Yukawa couplings of light quarks, but that there is no further 
quark-mass suppression. 
This is at  variance 
with   what is found in the standard  perturbative calculations of the light-quark contributions. 
We demonstrate   that  
the non-perturbative corrections modify the   $H \to \gamma Z$ 
and $H \to \gamma \gamma$ decay rates  by  
${\cal O}(10^{-5})$,
 well below
the expected precision with which such decays can be studied both 
at the high-luminosity LHC and at  future colliders. 
 }

\keywords{Higgs decays, non-perturbative physics}

\allowdisplaybreaks

\begin{document}
\maketitle

\section{Introduction}

Measurements of the Higgs boson couplings to electroweak gauge bosons are expected to reach  the percent-level precision at the high-luminosity LHC \cite{deBlas:2019rxi,Bass:2021acr}.  This  precision  
can be further improved at the future 
$e^+e^-$ colliders \cite{deBlas:2019rxi,Bass:2021acr}.  Since perturbative computations in the Standard Model,  commensurate with  this level of precision, are  feasible \cite{Actis:2008ts,Inoue:1994qj,Bonciani:2015eua,Degrassi:2004mx,Degrassi:2005ik,Djouadi:2005gi,Corrections:1996,Maierhofer:2012bt,Spira:1997fn,Spira:2001sd,Passarino:2006pm,Bardin:1990ws,GiuGiudice:2002ap,Djouadi:2005ik},    an unambiguous and 
instructive 
comparison of the results of the experimental measurements and the theoretical predictions is 
expected to be possible. 

This point of view, widely shared among particle physicists,   tacitly 
assumes that all  contributions to these decays, originating  from 
the \emph{non-perturbative} domain of QCD, are small, compared 
to the  percent or per mille accuracy.
This assumption  is very natural, as there is a  huge disparity between the mass 
of the Higgs boson $m_H$ and the energy scale of non-perturbative QCD 
$\Lambda_{\rm QCD}$. It is certain that the decays of  Higgs bosons to electroweak vector bosons and photons are predominantly determined by the 
short-distance physics.  
In contrast, the non-perturbative QCD effects involve large distances. For this reason they have   always  been thought to be negligible. 


Recently, this story received an interesting 
twist 
\cite{Knecht:2025nyo,Haisch:2025bat}. It has been  known for a 
long time  that the   contributions 
of light quarks 
to the $H \to \gamma \gamma$ decay
rate are proportional to the  \emph{second} power of their masses  ${\cal O}(m_q^2/m_H^2)$.   One   power of $m_q$  comes 
from the Yukawa coupling, and the other one   from the helicity flip on the 
light-quark line as  required for the decay of a scalar particle to two spin-one particles through a fermion loop. 

While the appearance of the relevant Yukawa coupling in any amplitude where the Higgs boson couples to a light quark  
is indisputable, it is less obvious that   the second power of the  light-quark mass is actually present  when   the \emph{non-perturbative} contribution to the Higgs decay amplitude is considered \cite{Knecht:2025nyo}. 
Indeed, it was argued in 
Ref.~\cite{Knecht:2025nyo}  that in the non-perturbative contribution 
this additional 
power of the quark mass is replaced by the quantity related to  the so-called  quark condensate
\cite{Gell-Mann:1968hlm}. Since the quark condensate density is the 
order parameter  of the spontaneous chiral 
symmetry breaking in  QCD \cite{Gell-Mann:1968hlm}, it remains non-vanishing 
in the massless quark limit,  eliminating 
the second power of the light quark mass in the Higgs decay amplitudes to electroweak vector bosons and photons.

 It was also argued in Ref.~\cite{Knecht:2025nyo}  that the non-perturbative effects in the decays 
 $H \to \gamma \gamma$ and $H \to  \gamma Z$ are likely to be enhanced 
 and, in the case of $H \to \gamma \gamma$, could be 
 as large as a few percent.  If these results  are  confirmed, 
 they will significantly affect the program of Higgs precision 
 studies at the high-luminosity LHC and at the future colliders,  since this  would mark the very first time  that   ${\cal O}(1\%)$ contribution to the $H \to \gamma \gamma $ decay is identified, that 
 cannot be fully controlled.
 
 More recently, 
  the non-perturbative QCD corrections to  the   $H \to \gamma \gamma$ 
  decay 
 were studied  in  Ref.~\cite{Haisch:2025bat}. There, the dispersion relation 
 in the invariant mass of  the two photons was  considered for the corresponding form factor.
  The spectral density in 
 the dispersion relation receives contributions from low-energy hadronic states, including 
 $\pi^+ \pi^-, K^+ K^-$ etc. 
 Estimating such \emph{hadronic} contributions 
 to the $H \to \gamma \gamma$ form factor,  
 the non-perturbative effects were found to be much smaller than the result in Ref.~\cite{Knecht:2025nyo} and, in fact, in line 
 with widely shared expectations that such effects can only appear 
 well below the percent-level precision target. 
 However, no 
 statement about the light-quark mass dependence was provided
 in Ref.~\cite{Haisch:2025bat},  probably because    the computational method used there is not conducive  to  addressing such a  question. 

Our goal in this paper is to discuss  the non-perturbative  corrections  to Higgs boson decays to electroweak vector bosons and photons  one more time.  In doing 
that, we  would like to {\it i}) elaborate on the issue 
of the degree of suppression 
of these decays by the 
light-quark masses,   and {\it ii}) provide alternative 
estimates of  the  magnitude
of the non-perturbative effects. 

We find it convenient 
to discuss the Higgs decay processes in a particular 
order, starting with  the $H \to Z^*Z^*$ transition,  
continuing with  
$H \to\gamma Z$ and, finally, arriving at  the $H \to \gamma \gamma$ decay.  This order is motivated by a degree of theoretical sophistication  
required  to perform 
the analysis of 
the non-perturbative effects in
a particular Higgs decay.  Indeed, while the non-perturbative phenomena in the $H \to Z^*Z^*$ transition can be analyzed within the framework of the short-distance 
operator product expansion 
\cite{Wilson:1969zs},
description of the 
$H \to \gamma Z$ and $H \to \gamma \gamma$ decays  
requires an operator product expansion on the light cone
\cite{Brodsky:1981kj,Efremov:1979qk,Chernyak:1983ej,Balitsky:1989ry,Braun:1988qv}
which introduces non-perturbative quantities beyond the familiar vacuum condensates of quark and gluon fields. 

In general, methods that we employ in this analysis 
are closer to the approach described in Ref.~\cite{Knecht:2025nyo},  and we will confirm the absence of the 
second power of the light-quark mass in the non-perturbative corrections  to the relevant Higgs decay amplitudes.  However,  our  numerical estimate of the non-perturbative 
contributions in $H \to \gamma \gamma$ are a factor 
$10^{-4}$ \emph{smaller} than the results in Ref.~\cite{Knecht:2025nyo}
and, thus, are fully  in line  with those reported in  Ref.~\cite{Haisch:2025bat}.

\section{Higgs boson transition to two Z bosons}
\label{sect2}

There is a direct coupling of the Higgs boson to two $Z$ bosons in the Standard Model, $ m_Z^2/v \; H Z_\mu Z^\mu$,
where $m_Z$ is the $Z$-boson mass and $v$ is the Higgs-field vacuum expectation value. This coupling does not lead to a decay to two on-shell $Z$-bosons because 
$m_H < 2m_Z$. The decay 
$H \to Z Z^*$, where  one of the $Z$ bosons is 
off-shell, has a ${\cal O}(3\%)$ branching ratio in the Standard Model. 
\begin{figure}
\begin{center}
\includegraphics[scale=1.0]{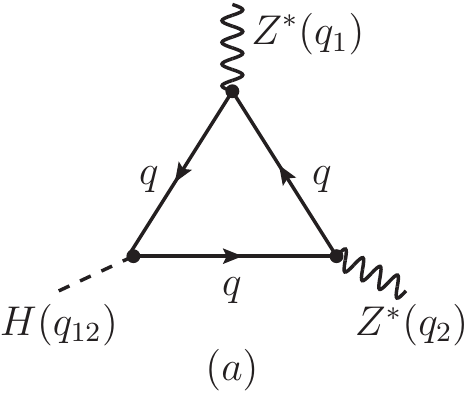}
\hspace{1cm}
\includegraphics[scale=1.0]{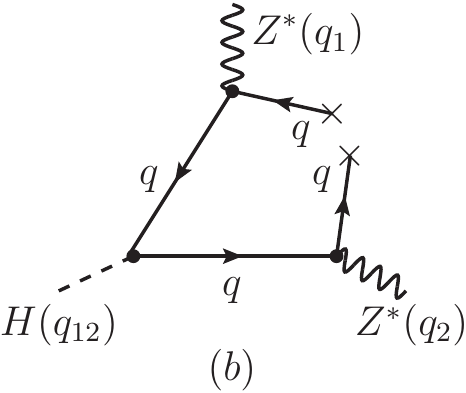}
\end{center}
\caption{Contribution of light quarks to the  $H \to Z^*Z^*$ transition: (a) an example of a short-distance perturbative  diagram, (b) an example of the  non-perturbative contribution.  Terminated 
quark lines imply the quark condensate. 
}
\label{fig:diag}
\end{figure}
There are several ways in which the 
$H \to Z Z^*$ decay can be affected by the non-perturbative QCD effects, and we will not discuss all of them. 
In fact, we are particularly interested in the non-perturbative effects related to the light-quark contributions   to the  $H \to Z Z^*$  transition,  
because  the same  effects will  provide 
the \emph{leading}  non-perturbative corrections to the  $H \to \gamma Z$ and 
$H \to \gamma \gamma$ decays. 

To understand why these light-quark contributions are peculiar, we write the amplitude of the  $H(q_{12}) \to Z^*(q_1) + Z^*(q_2)$ transition,  moderated by a 
light  quark $q$, in the following way (see 
Fig.~\ref{fig:diag})
\be
{\cal M}^{q}_{H\to ZZ} = y_q \; g_{Zq}^2 \; T^{\mu \nu} \;\epsilon^{(1)}_{\mu} \epsilon^{(2)}_{\nu}.
\label{eq2.1}
\ee
In Eq.~(\ref{eq2.1}) $y_q=m_q/v$ is the light-quark Yukawa coupling,
$g_{Zq}$ is the vector coupling constant between the 
$Z$ boson and the quark $q$, and  $\epsilon^{(i)}$, $i=1,2$,  
are the polarization vectors of the two $Z$ bosons that 
satisfy the standard transversality  conditions,  $\epsilon^{(i)} \cdot q_i 
= 0$.
Furthermore, 
\be
T^{\mu\nu}=\langle 0| \;\hat T^{\mu\nu}\; |0\rangle\,,
\label{eq2.2}
\ee
where the operator $\hat T^{\mu\nu}$ is the time-ordered product of three currents,
\be
\hat T^{\mu\nu}=\!\int\! d^4x \,d^4 y\, e^{iq_1\cdot x+iq_2\cdot y} \,T\left\{\bar\psi_q(0)\psi_q(0), \,
\bar\psi_q(x)\gamma^\mu\psi_q(x),\, \bar\psi_q(y)\gamma^\nu\psi_q(y)\right\}\,.
\label{eq2.3}
\ee
In Eq.~(\ref{eq2.3})  $\psi_q$ is the field operator of  the quark
$q$, and for each current the summation over quark colors is assumed. 
We have dropped 
the axial-vector coupling of the $Z$-bosons to light 
quarks for simplicity, and because keeping it does not  affect our reasoning 
and conclusions. 
For 
the on-shell $Z$ bosons, $q_1^2 = q_2^2 = m_Z^2$; 
however,  it is more convenient to keep these invariant masses as free parameters in the computation.

In perturbation  theory 
\be
T^{\mu \nu} = -i N_c
\left \langle 
\gamma^\mu \frac{1}{\slashed k  - \mq } \,\gamma^\nu \frac{1}{\slashed k + \slashed q_2 - \mq} \,\,
\frac{1}{\slashed k - \slashed q_1  - \mq} 
\right \rangle + 
( \mu \leftrightarrow \nu,\, q_1 \leftrightarrow q_2),
\label{eq2.4}
\ee
where 
\be 
\langle \cdots \rangle 
= \int \frac{{\rm d}^4 k}{(2\pi)^4}
{\rm Tr} \left [ \cdots \right ].
\ee

It is easy to see  that 
the amplitude $T^{\mu \nu}$ 
in Eq.~(\ref{eq2.4}) 
vanishes if  $\mq$ is set to zero, i.e. in the chiral limit.  Hence,
we conclude that perturbatively $T^{\mu \nu} \sim m_q$\,.\footnote{\,For clarity, we note that we 
do not discuss here the so-called singlet contribution to Higgs decays, that proceeds through an intermediate $gg$ state, i.e. $H \to gg \to ZZ$. Singlet corrections  receive contributions from the light-quark loops,  but they  are not proportional to the light-quark masses. However, they only appear in higher perturbative orders.  } Since 
the Yukawa coupling  $y_q$ in Eq.~(\ref{eq2.1}) is also 
proportional to $\mq$, we find ${\cal M}^q \sim m_q^2$ which is  
the familiar quadratic dependence  of the $H \to ZZ^*$
decay amplitude  facilitated by  the  light quark $q$. We note, that the integral over the loop momentum 
$k$ in the perturbative amplitude is saturated at  $k \sim q_1 \sim q_2\gg m_q$\,.

However, in the integral  over the loop momentum $k$, there are regions where the perturbative expansion  breaks down. This happens whenever the momenta  comparable to  the hadronic scale $\Lambda_{\rm QCD} \sim 1~{\rm GeV}$ flow through quark propagators.  
To find the non-perturbative contributions to the amplitude that originate from these regions,  we apply the operator product expansion (OPE) \cite{Wilson:1969zs,Shifman:1978bx} to  products of  currents in the operator $\hat T^{\mu\nu}$  in  Eq.\,(\ref{eq2.3}),
\be
\hat T^{\mu\nu}=\sum_i C^{\mu\nu}_i(q_1, q_2) \,{\cal O}_i\;.
\label{eq2.5}
\ee
In the above equation, $C_i^{\mu \nu}$ and ${\cal O}_i$ are the  
Wilson coefficients and 
local operators, respectively. 
The leading operator associated with the non-perturbative 
regions is ${\cal O}_q=\bar \psi_q \psi_q$ where the summation over  colors is implied. The vacuum average of 
this operator is the quark condensate \cite{Gell-Mann:1968hlm}.

It is straightforward  to compute  the OPE coefficient $C^{\mu\nu}_q$ for the operator ${\cal O}_q$. To this end, it is sufficient to calculate the matrix element 
\be
\langle q(p_2)|\,\hat T^{\mu\nu}\,|q(p_1)\rangle
\label{eq2.6}
\ee
in the limit of vanishing momenta $p_{1,2}$ of  the initial and final quarks which 
we take to be \emph{massless}. Diagrammatically, this calculation involves 
cutting various propagators  in the perturbative triangle diagram (see Fig.~\ref{fig:diag}b)
and, assuming that  vanishingly-small momentum flows through the cut propagator,   computing the tree  amplitude composed of the product of the two remaining quark propagators through which the large momenta  $\sim q_{1},q_{2}$ flow. 

For simplicity, we write the resulting OPE coefficient of the $ q \bar q$ operator contracted with the polarization vectors of the two $Z$-bosons, 
\be
C_q\!=C_q^{\mu\nu}\epsilon^{(1)}_{\mu}\epsilon^{(2)}_{\nu}\!=-\frac{q_{12}^2}{q_1^2\,q_2^2}\left\{\!\big[(1\! -\! (r_1\!-\!r_2)^2\,\big]\epsilon^{(1)}\!\cdot\!\epsilon^{(2)} \!-2\big[1\! +\!r_1+\!r_2\,\big]
\frac{q_2\cdot\!\epsilon^{(1)}\;q_1\!\cdot\epsilon^{(2)}}{q_{12}^2} \!\right \}\!.
\label{eq2.7}
\ee
We have defined  $r_i=q_i^2/q_{12}^2$,\; $i=1,2$, and have used the transversality 
conditions $\epsilon^{(i)} \cdot q_i=0$,
$i=1,2$, to simplify the above equation.

Using this result to find the non-perturbative contribution of one nearly  massless  quark to the 
Higgs decay amplitude in Eq.~(\ref{eq2.1}), we obtain 
\be
{\cal M}_{H \to ZZ}^{q,\,\rm np} = y_q g_{Zq}^2 \, 
\langle 0 | \bar \psi_q \psi_q | 0 \rangle\,
C_q\,.
\label{eq2.8}
\ee

To estimate the numerical impact of this contribution, we note that the non-perturbative amplitude in Eq.~(\ref{eq2.8}) interferes   with  the leading order amplitude of   the  $H \to Z^*Z^*$ transition, ${\cal M}_{H \to ZZ} =2( m_Z^2/v )\,\epsilon^{(1)}\!\cdot \!\epsilon^{(2)}$.   Thus, the correction is proportional to  the ratio of non-perturbative and perturbative amplitudes, which  evaluates to
\be
\frac{{\cal M}_{H \to ZZ}^{q,\,\rm np}}{{\cal M}_{H \to ZZ}}
\approx -(2 \pi \alpha) \; \frac{m_q \langle 0|\bar \psi_q \psi_q |0\rangle }{m_Z^4}. 
\label{eq2.9a}
\ee
To obtain the above result, we used $y_q = \mq/v$, 
$g_{Zq}^2 \sim 4 \pi \alpha$, 
${\cal M}_{H \to ZZ} \sim 2 m_Z^2/v$ and $C_q\sim -m_Z^{-2}$, since for the purpose 
of this order-of-magnitude estimate, we take $m_H^2 = q_{12}^2 \sim q_1^2 \sim q_2^2 \sim m_Z^2$.

To obtain the full result, one must  sum over all light-quark flavors. This sum is strongly dominated by the strange-quark contribution. 
Hence, focusing on the strange quarks   and using the  
Gell-Mann-Oakes-Renner relation \cite{Gell-Mann:1968hlm},
we write\footnote{\,In Eq.~(\ref{eq2.9a}), $m_q$ and 
$\langle 0| \bar \psi_q \psi_q 
| 0 \rangle$ are defined at 
the high renormalization scale 
$\mu \sim m_Z$. However, since 
the product of the quark mass 
$m_q$ and the quark condensate $\langle 0| \bar \psi_q \psi_q | 0 \rangle$ 
does not depend on the renormalization scale, 
we can use the Gell-Mann-Oakes-Renner relation, naturally associated with low hadronic scales,  to estimate 
their product.} 
\be
\frac{{\cal M}_{H \to ZZ}^{\rm np}}{{\cal M}_{H \to ZZ}}\approx  
\frac{(\pi \alpha) f_K^2 m_K^2}{m_Z^4} \sim {\rm few}  \times 10^{-12},
\label{eq2.9b}
\ee
where we used  
$f_K = 155.7~{\rm MeV}$ \cite{FLAG:2024oxs} and 
$m_K = 498~{\rm MeV}$.

We note that  non-perturbative corrections to the  
$H \to W^+W^-$ transition  can be estimated along the same lines.  Because in this case 
flavor-changing quark currents are involved, 
the details of the analysis will be different, 
but the numerical suppression will be similar 
to the result in Eq.~(\ref{eq2.9b}).

Hence,  in spite of being proportional to 
the first power of the 
light-quark mass only, 
the non-perturbative correction  
to $H \to Z Z^*$ appears to be  tiny. There are two reasons for  this very strong suppression. The first one is the   fine-structure constant  $\alpha$, which appears 
because we compute the 
loop-induced non-perturbative 
correction  to the decay amplitude $H \to Z Z^*$ which  by  itself  is not  loop-induced.

The second reason for  the suppression is the fourth power of the hard scale $q_{1,2} \sim m_Z$ in the denominator in Eq.~(\ref{eq2.9a}).
This high power appears 
because the mass-dimension of the 
non-perturbative 
matrix element $\langle 0 | m_q \bar \psi_q \psi_q |0 \rangle$  is equal to four. 
Since there are other non-perturbative quantities of the same mass-dimension, for example, the gluon condensate $\langle 0|\alpha_s/\pi \; G_{\mu \nu} G^{\mu \nu} | 0 \rangle$, there are  other non-perturbative corrections to the $H \to ZZ^*$ transition  that are of the same order 
as the light-quark contribution shown 
in Eqs~(\ref{eq2.9a}),\,(\ref{eq2.9b}). Hence, 
the light-quark 
contribution to ${\cal M}^{\rm np}_{H \to ZZ}$ is  certainly peculiar but not unique in any way.

We will discuss the non-perturbative contributions 
to $H \to \gamma Z$ and to $H \to \gamma \gamma$ in the following sections, and
it is interesting that in those cases 
 \emph{both of the above points 
become invalid}.  Indeed,  because  the  $H \to \gamma Z$ and $H \to \gamma \gamma$ decays 
are both loop-induced, the fine-structure constant  or any other electroweak coupling constant will not 
be present in the ratios of non-perturbative and perturbative amplitudes in these  cases. 

We will also see  
that the non-perturbative contributions 
of light quarks to $H \to \gamma Z$ 
and $H \to \gamma \gamma$ are suppressed 
by the  \emph{second}  power  of the hard scale and, therefore, are much larger.  One can anticipate this  because   the Wilson coefficient in Eq.~(\ref{eq2.7})  
reads  $C_{q} \sim q_{12}^2/(q_1^2q_2^2)$.
Hence, a naive  extrapolation of the above result to regions where  $q_{1,2}^2 = 0$,  which is exactly what is needed to describe the Higgs boson decays $H \to \gamma Z$ and $H \to \gamma \gamma$, indicates that a significant enhancement of the 
non-perturbative effects in such decays can be expected \cite{Knecht:2025nyo}. 

As we explain in the next section, 
this expectation is partially correct. Technically, to arrive 
at this result, we need to appreciate that 
the operator product expansion for processes  with 
a  massless final-state particle becomes different. 
In fact,  the proper theoretical framework to analyze the  $H \to \gamma Z$ decay is  an operator product expansion near the light cone, familiar from studies of hard 
exclusive processes 
\cite{Chernyak:1983ej,Brodsky:1981kj,Efremov:1979qk,Ball:2002ps,Braun:1988qv,Balitsky:1989ry}.

 \section{Higgs boson decay to photon and Z boson}

Consider the Higgs boson decay to a  photon with the momentum 
$q_1$ and a  $Z$-boson with the momentum $q_2$, $H(q_{12}) \to \gamma(q_1) + Z(q_2)$.
The amplitude for this process reads
\be
{\cal M}_{H \to \gamma Z} = y_q\, e_q\, g_{\,Zq} \; 
T^{\mu \nu} \epsilon^{(1)}_\mu \epsilon^{(2)}_\nu, 
\label{eq3.1}
 \ee
where $T_{\mu \nu}$ 
can be taken from Eqs~(\ref{eq2.2}),\,(\ref{eq2.3}) and 
we should set 
$q_1^2 = 0$ there. Furthermore, $e_q$ is the electric charge 
of the quark $q$. 
\begin{figure}
\begin{center}
\includegraphics[scale=1.0]{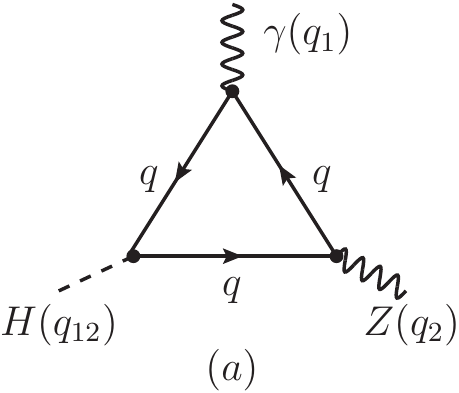}
\hspace{2cm}
\includegraphics[scale=1.0]{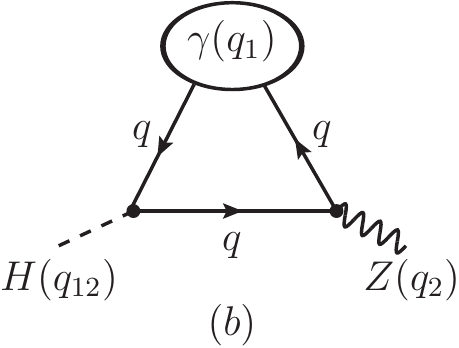}
\end{center}
\caption{(a) 
The perturbative short-distance contribution 
to $H \to Z \gamma$.
(b)\,
The non-perturbative fragmentation of a $q \bar q$ pair to the photon distribution amplitude.
 Diagrams with the 
 opposite fermion-flow direction are not shown.
}
\label{fig:diagL}
\end{figure}
Leading contributions to the 
$H \to  \gamma Z$ decay 
amplitude  are shown in 
 Fig.~\ref{fig:diagL}. The first, shown 
 in Fig.~\ref{fig:diagL}a, is entirely short-distance one and can be computed in perturbation theory.  Similarly to the case  $H \to ZZ^*$, it vanishes in the chiral limit, and we do not 
 discuss it further. The second contribution shown  in Fig.~{\ref{fig:diagL}(b) is more complex as it involves both the  short- and the  long-distance parts. The short-distance part 
 is  the transition of the 
 Higgs boson to the $Z$-boson and a collinear 
 $q \bar q $ pair. The  long-distance 
 part corresponds to the fragmentation 
 of a nearly collinear $q \bar q $ pair 
 to a photon. 
 
 The $ q \bar q \to \gamma $ fragmentation is  described by the so-called 
 photon distribution amplitude which was
first introduced in 
Ref.~\cite{Balitsky:1986st}, (see also Ref.~\cite{Balitsky:1989ry})
in the context of QCD sum rules
based on the light-cone operator product expansion. In 
  Ref.~\cite{Ball:2002ps} a comprehensive discussion  of these distribution amplitudes, including 
the separation of short- and long-distance effects in  the photon-quark
interactions can be found.

In principle, without further ado, this standard methodology\,\footnote{\,For example, in Refs~\cite{Khodjamirian:1995uc,Ali:1995uy}
light-cone sum rules with the photon distribution amplitude have been used  to describe the 
$B\to \gamma \ell\nu_\ell$ decay and a correlation function similar to 
Eq.~(\ref{eq3.2}), albeit  with completely different quark currents, was computed. }
\,can be straightforwardly applied to analyze 
the non-perturbative contributions to the 
$H \to \gamma Z$ decay.  However, 
we find it useful to
discuss the main ideas behind this approach.  To this end, we note that  the 
long-distance fragmentation phenomenon 
that we need to describe makes it inconvenient to work with the 
correlator $ \hat T_{\mu \nu}$. 
Instead, we  write $e_q T^{\mu\nu}\epsilon^{(1)}_\mu$ as 
the  matrix element of the $T$-product of the scalar and vector currents 
between the vacuum and the single-photon state 
\be
\begin{split}
&{\cal M}_{H \to \gamma\, Z} = y_q \, g_{\,Zq} \; 
\langle \gamma(q_1)\,|\hat\Pi^{\nu}\,|0\rangle \,\epsilon^{(2)}_\nu, \\
&\hat\Pi^{\nu}=i\int d^4x\, e^{-iq_{12} \cdot x}\, T\big\{
  \bar{\psi_q}(x)\psi_q(x),\bar{\psi_q}(0)\gamma^\nu \psi_q(0)\big\}\,.
\end{split}
\label{eq3.2}
 \ee
Equation (\ref{eq3.2}) contains both, perturbative 
and non-perturbative, contributions 
shown in Fig.~\ref{fig:diagL}.  However, 
for the massless quarks the perturbative 
contribution vanishes. Then, 
Eq.~(\ref{eq3.2}) is very convenient since 
the non-perturbative long-distance 
physics is isolated into  
the matrix element,  and 
the short-distance physics  is described by   the product of two current 
operators that appear 
explicitly in ${\hat \Pi}^{\nu}$.

To construct the OPE of  the product of currents in $\hat\Pi^\nu$ at small $x$, 
we compute  
its matrix element between the on-shell massless quark states with momenta $p$ and $p+q_1$.  Since $q_1^2=0$, the requirement 
that $(p+q_1)^2 = 0$ and $p^2 = 0$ implies 
that $p \cdot q_1 = 0$, which means 
that $p$ is either aligned with $q_1$ or 
is transversal to it. 
At tree level, we find 
\be
\langle q(p+q_1)|\,\hat\Pi^\nu\,|q(p)\rangle=-\bar u_i(p+q_1)
\left[\frac{1}{\slashed p -\slashed q_2}\gamma^\nu +
\gamma^\nu \frac{1}{\slashed p+\slashed q_{12}}\right]u^i(p)\,,
\label{eq3.3}
\ee
where $u^i(p)$ is the spinor wave function for the quark with the color $i$ and momentum $p$\,. 
Rewriting products of Dirac matrices 
through anti-commutators and  
commutators, we obtain 
\be
\begin{split}
\langle q(p+q_1)|\,\hat\Pi^\nu\,|q(p)\rangle=&-
\left[\frac{(p-q_2)^\nu}{q_2^2 -2pq_2} + \frac{(p+q_{12})^\nu}{q_{12}^2 +2pq_{12}} \right]\bar u_i(p+q_1)u^i(p)\\
&-\left[\frac{(p-q_2)_\alpha}{q_2^2 -2pq_2} - \frac{(p+q_{12})_\alpha}{q_{12}^2 +2pq_{12}}\right]\bar u_i(p+q_1)\sigma^{\alpha\nu} u^i(p)\,, 
\end{split}
\label{eq3.4}
\ee
where $\sigma^{\alpha \nu} = [\gamma^\alpha, \gamma^\nu]/2$.

It is convenient to start 
with the case of the small photon and quark momenta, $p,\,q_1 \ll q_2$ which corresponds 
to the limit when the Higgs and the $Z$ boson masses are very close. 
In this limit, 
the first term in Eq.~(\ref{eq3.4}) drops 
and the second survives. Rewriting 
$\bar u \sigma^{\alpha \nu} u$ in terms 
of the quark-field operators, we conclude  that  the OPE of $\hat \Pi^\nu$ takes the form 
\be
\hat\Pi^\nu=\frac{2q_{2\alpha}}{q_{2}^2}\, \; \bar \psi_q(0) \,\sigma^{\alpha\nu} \psi_q(0)\,.
\label{eq3.5}
\ee
It remains to compute the matrix element 
of ${\hat \Pi}^\nu$ in 
Eq.~(\ref{eq3.5}) 
between the photon and  vacuum states. 
This matrix element 
is known \cite{Ioffe:1983ju}; it 
is parametrized by a 
particular quantity $\chi$ called the magnetic 
susceptibility of the quark condensate,
\be
\langle \gamma(q_1)|\bar{\psi_q}(0)\sigma_{\alpha\nu} \psi_q(0)|0\rangle=
e_q\,\chi\,  \,\langle0|\bar{\psi_q}\psi_q |0\rangle f^{(1)}_{\alpha\nu}\,.
\label{eq3.6}
\ee
In this equation, $\langle 0|\bar{\psi_q}\psi_q |0\rangle$ is the quark condensate, and 
$f^{(1)}_{\alpha\nu}=q_{1\,\alpha}\epsilon^{(1)}_{\,\nu}-q_{1\,\nu}\epsilon^{(1)}_{\,\alpha}$ is the field-strength tensor of the 
photon.  Thus, we conclude that the non-perturbative part of $H\to \gamma Z$ amplitude in the limit of small photon 
momentum $q_1$, i.e. for $m_H^2 - m_Z^2 \ll m_H^2$, reads 
\be
{\cal M}_{H \to \gamma Z}^{q,\;{\rm np}} \; =\;  e_q \,g_{Zq}\, y_q
\; \frac{\langle 0| \bar \psi_q \psi_q |0\rangle\ \chi}{m_H^2}
\;
f^{(2)}_{\mu \nu} f^{(1){\mu \nu}} \,,
\label{eq3.7}
\ee
where $f^{(2)}_{\mu \nu} = q_{2\,\mu} \epsilon^{(2)}_{\nu} 
- q_{2\,\nu} \epsilon^{(2)}_{\mu} $.
An interesting property  of this result is that the non-perturbative 
amplitude  is only 
suppressed by the \emph{second} power of the hard scale. This feature is related to the appearance 
of the magnetic susceptibility $\chi$ which 
has the mass-dimension $-2$ and,   parametrically, is determined by the soft QCD scale, $\chi \sim {\cal O}(\Lambda_{\rm QCD}^{-2})$. Thus, it provides 
an enhancement of the non-perturbative effects in $H \to \gamma Z$ decay,    which 
was advertised at the end of the previous  section, albeit so far derived only for the   
unphysical case $m_H \approx  m_Z$.

We continue with the discussion of  the realistic 
 case, where  the photon momentum is of the same order as $m_H$ and $m_Z$.  To this end, we expand  Eq.\,(\ref{eq3.4}) to higher powers in the quark momentum $p$ to 
 derive the OPE coefficients of  operators of higher mass-dimensions.\footnote{\,Since we are interested 
 in  the matrix of the operator ${\hat \Pi}^\mu$
 between the photon and the vacuum, we can discard  the 
 first term on the right-hand side of Eq.~(\ref{eq3.4}).
 } 
The leading operator is $\bar \psi_q\,\sigma^{\alpha\nu} \psi_q$, 
that was already introduced; it has   mass-dimension $3$ and spin  $1$, so its twist\,\footnote{\, Twist of an operator 
is the difference between its mass-dimension and  spin.} 
is $2$. Powers of momentum $p$ in the expansion  would lead to the appearance of higher spin operators with derivatives of the quarks fields. For example, a term that is linear in $p$ introduces the following  operator 
\be
{\cal O}_\mu^{\,\alpha\nu}=\bar \psi_q\,\sigma^{\alpha\nu} iD_\mu \psi_q\,,
\label{eq:oper}
\ee
where $D_\mu$ is the covariant derivative. 
It  provides the following  addition  to the leading contribution to $\hat \Pi^\nu$ in  Eq.~(\ref{eq3.5}) 
\be
\delta  \,\hat \Pi^\nu \sim \frac{q_2^\mu q_{2\,\alpha} }{(q_2^2)^2} 
\; 
{\cal O}_\mu^{\alpha \nu}.
\ee
Normally, $\delta {\hat \Pi}^\nu$ is 
a small correction to ${\hat \Pi}^\nu$,
but since  we are interested in the matrix element of 
${\hat \Pi}^\nu $ between the photon with a large momentum 
$q_1$ and the vacuum, this is not 
true anymore. Indeed, the relevant matrix element is proportional to the photon momentum 
\be
\langle  \gamma(q_1) | \; {\cal O}_\mu^{\alpha \nu}  \; 
| 0 \rangle \sim q_1^\mu  \; \langle  \gamma(q_1) |  
\bar \psi_q \sigma^{\alpha \nu } \psi_q \; | 0 \rangle, 
\ee
which 
implies that for $q_1 \cdot q_2 \sim q_2^2$,
\be
\langle  \gamma(q_1)\;  | \; \delta \hat \Pi^\nu \; |
0 \rangle \sim 
\langle \gamma(q_1) \;  | \; \hat \Pi^\nu \; | 0 \rangle,
\ee
and there is no suppression. 
Hence,  all  terms with additional derivatives acting 
along the light-cone direction, defined  by the photon 
momentum $q_1$, cannot be discarded.  The summation of all such 
contributions provides  a non-perturbative  object that is known as the 
twist-two photon distribution amplitude \cite{Balitsky:1989ry}. 

The twist-two photon distribution amplitude depends 
on the ratio of the  hard scale of the process we are interested in, 
and the non-perturbative QCD scale $\Lambda_{\rm QCD}$.  In our case, this ratio is very large 
$m_H/\Lambda_{\rm QCD} \sim 10^3$.
Because of this, 
we are  interested 
in the so-called asymptotic form of this amplitude
\cite{Chernyak:1983ej} which 
is obtained by taking the hard scale to infinity. To introduce  it, we note that 
the  operator ${\cal O}_\mu^{\alpha \nu}$
can be re-written in the following way  
\be
{\cal O}_\mu^{\,\alpha\nu}=\frac{1}{2} \,i\partial_\mu\big(\bar \psi_q\,\sigma^{\alpha\nu} \psi_q\big) + \frac{1}{2}\,\bar \psi_q\,\sigma^{\alpha\nu} i\overset{\leftrightarrow}{D}_\mu \psi_q\,.
\label{eq:oper1}
\ee
The first term  in 
the above equation 
is the total derivative of the leading operator whose matrix element in Eq.\,(\ref{eq3.6}) defines the magnetic susceptibility.
The matrix element of the second term in Eq.\,(\ref{eq:oper1}) 
has a similar form
but it is a different operator nonetheless.
 In principle, one should define its  matrix element by introducing another susceptibility-like quantity  that 
will differ 
from the susceptibility $\chi$ in Eq.\,(\ref{eq3.6}). 

In principle, 
the contribution of \emph{both} operators in Eq.\,(\ref{eq:oper1}),
as well as   all  other 
multi-derivative operators,  must be taken into account. However,  
it is known \cite{Chernyak:1983ej} that 
all operators which  in addition to 
total derivatives, 
contain other quantities, are 
suppressed by the  
logarithm
$\log(m_H/\Lambda_{\rm QCD})$ which,  as we already mentioned,  is large. 
Hence, in the limit 
$m_H \gg \Lambda_{\rm QCD}$ only operators 
that are total 
derivatives 
of $\bar \psi_q \sigma^{\alpha \nu} \psi_q$ should be retained.  
These operators provide the asymptotic form of the photon distribution 
amplitude  which is  therefore  completely determined by the  single non-perturbative parameter $\chi$.

The photon distribution amplitude 
$\phi_\gamma(\xi)$ describes how the photon 
momentum $q_1$ is shared between a quark and an antiquark fragmenting into the  photon. Our convention is that  the antiquark  carries   
momentum $\xi q_1$ and the quark carries the rest. 
To account for this, we write $p^\mu = -\xi q_1^\mu$ in Eq.\,(\ref{eq3.4}), extending the  operator $\hat \Pi^\nu$ to non-vanishing quark momenta
\be
\hat\Pi^\nu=
\int \limits_{0}^{1} {\rm d} \xi \; \phi_\gamma(\xi) \;
\left [ 
  \frac{1}{(1-\xi)  q_{12}^2+ \xi q_2^2}
  + 
  \frac{1}{\xi   q_{12}^2+ 
  (1-\xi)  q_2^2}
  \right ] q_{2\,\alpha}\bar \psi_q\,\sigma^{\alpha\nu} \psi_q\, .
\label{eq3.8}  
\ee
 Finally, taking the matrix element
between the photon and the vacuum 
state, making 
use of Eq.~(\ref{eq3.6}) 
and substituting  $q_{12}^2 = m_H^2, 
\; q_2^2 = m_Z^2$, we find 
\be
{\cal M}_{H \to \gamma Z}^{q,\rm np} 
= \;  e_q g_{Zq} y_q
\; \frac{\langle \bar \psi_q \psi_q \rangle \chi}{2 m_H^2}
\;
f^{(2)}_{\mu \nu} f^{(1)\,{\mu \nu}}  \!\int \limits_{0}^{1} 
{\rm d} \xi 
\left [ 
  \frac{1}{1  -(1-r) \xi}
  + 
  \frac{1}{r + \xi (1-r)}  
  \right ] 
 \phi_\gamma(\xi).
 \ee
 where $r = m_Z^2/m_H^2$. To compute 
 the remaining integral, we employ 
 the asymptotic form of the 
 photon distribution amplitude\,\footnote{ 
\,The asymptotic form of the 
\emph{pion} distribution amplitude 
is derived in Ref.~\cite{Chernyak:1983ej}, 
but it can be equally well  
applied to the photon case.}
 \be
\phi_\gamma = 6\, \xi (1-\xi), 
 \ee
and obtain 
 \be
{\cal M}_{H \to \gamma Z}^{q,\rm np} =e_q \, g_{Z q} \; y_q
\; \frac{\langle \bar \psi_q \psi_q \rangle \chi}{m_H^2}\;
f^{(2)}_{\mu \nu} f^{(1)\,{\mu \nu}} 
\; 3F\left (\frac{m_Z^2}{m_H^2}\right ),
\label{eq3.10}
\ee
where 
\be
F(r) = \frac{1+r}{(1-r)^2} + \frac{2r}{(1-r)^3} \ln r\;.
\label{eq3.11}
\ee
The function $F(r)$ is finite at $r = 0$ and at $r=1$,
and for $r \in [0,1]$ assumes numerical values between  $1$  and  $1/3$.
To estimate the numerical impact of the non-perturbative corrections to  the  $H \to \gamma Z$ decay amplitude on the  decay rate, we note that 
the main effect comes from the interference of ${\cal M}_{H \to \gamma Z}^{\rm np}$ with 
the leading perturbative  amplitude  that contains loops of heavy quarks and vector bosons. 
This amplitude was computed in Ref.~\cite{Lee:1974hq} for the first time. 

Given the fact that the non-perturbative effects that we discuss in this paper  are quite   small,  it is  sufficient to  provide a rough estimate  
of the perturbative amplitude. 
To this end, we write  
\be
{\cal M}_{H\to\gamma Z} \sim  \frac{
e \; g_{Z}
}{4 \pi v} \;  f^{(2)}_{\mu \nu} f^{(1)\mu \nu},
\label{eq3.18a}
\ee
where $g_Z$ is the electroweak coupling constant. 
The ratio of the non-perturbative and perturbative amplitudes evaluates to 
\be
\frac{ {\cal M}_{H \to \gamma Z}^{q,\rm np}}{{\cal M}_{H \to \gamma Z}} \sim \frac{6 \pi  \; m_q \; Q_q  \chi \langle \bar \psi_q \psi_q \rangle }{m_H^2} 
\to 
 -\frac{3 \pi\; Q_s  \chi f_K^2 m_K^2 }{m_H^2}\,, 
 \label{eq3.13}
\ee
where we have used $F(m_Z^2/m_H^2) \approx 1/2$ for the physical masses of the  $Z$ and Higgs bosons. Furthermore,  in the last step we took 
into account that strange quarks provide the largest contribution and again 
used the Gell-Mann-Oakes-Renner 
relation \cite{Gell-Mann:1968hlm}.

Similarly to the 
case $m_H \sim m_Z$ discussed earlier,  the most striking feature of 
Eq.~(\ref{eq3.13}) in comparison with the $H \to ZZ^*$ case, is that  the degree of suppression is reduced from 
$1/m_Z^4$, to $1/m_H^2 
\sim 1/m_Z^2$.
This (dimension-full) difference is accounted 
for by the magnetic susceptibility of the vacuum $\chi$, whose  mass-dimension is  minus two.  Writing  
$\chi = M_\chi^{-2}, 
$
we express  Eq.~(\ref{eq3.13}) 
as follows
\be
\frac{ {\cal M}_{H \to \gamma Z}^{\rm np}}{{\cal M}_{H \to \gamma Z}} \sim 
 -\frac{3 \pi\; Q_s   f_K^2 m_K^2 }{m_H^2 M_\chi^2}. 
\label{eq:ratioZZ}
\ee
For numerical estimates, we require magnetic  susceptibility $\chi$ at the high scale.  However, for 
simple numerical estimates, we 
will neglect its running. 
We  use $\chi(\mu = 1~{\rm GeV}) = 2.85 \pm 0.5~{\rm GeV}^{-2}$ \cite{Rohrwild:2007yt}, 
\footnote{\,Similar estimates  of the magnetic susceptibility
were obtained 
in Refs~
\cite{Ioffe:1983ju,Ball:2002ps,Bali:2012jv}.
} 
so that 
$M_\chi$ evaluates to ${\cal O}(0.6~{\rm GeV})$. This implies 
\be
\frac{ {\cal M}_{H \to \gamma Z}^{\rm np}}{{\cal M}_{H \to \gamma Z}} \sim 
{\rm few} \times 10^{-5}.
 \ee
Therefore, we find that, in the 
$H \to \gamma Z$ case, the non-perturbative effects are suppressed by only \emph{two} powers of the hard scale, whereas in the $H \to ZZ^*$ case, 
they are suppressed by  \emph{four} powers. 
This is in line  with the proposal 
in Ref.~\cite{Knecht:2025nyo},
which effectively advocates 
the replacement of both factors $1/q_i^2$, $i=1,2$,  in 
Eq.~(\ref{eq2.7}) with the factor $1/M_V^2$, where $M_V$ is a  mass of a typical light vector meson $M_\rho, M_\omega, M_\phi$, 
as a way to describe 
the non-perturbative contributions to 
$H \to  \gamma Z$ and, eventually, to  $H \to \gamma \gamma$ decays. 
In the next section, we will discuss whether  the extension of our analysis  to the $H \to \gamma \gamma $ case supports this approach.

\section{Higgs boson decay to 
two photons }

It remains to discuss the $H \to \gamma \gamma$ case. 
We can derive the non-perturbative corrections to the 
$H \to \gamma \gamma $ amplitude by  using the results for  the 
$H \to \gamma Z$ amplitude  
discussed in the previous section, 
and extrapolating them to $q_2^2 = 0$. 
This extrapolation is straightforward  because the function $F(r)$ possesses  smooth $r = q_2^2/m_H^2  \to 0$ limit, $F(0)=1$. Thus,
\be
{\cal M}_{H \to \gamma \gamma}^{q,\,\rm np}
=6\,e_q^2\, y_q
\; \frac{\langle \bar \psi_q \psi_q \rangle \,\chi}{m_H^2}\;
f^{(2)}_{\mu \nu} f^{(1)\,{\mu \nu}}\,.
\label{eq4.1}
\ee
In comparison to Eq.~(\ref{eq3.10}),  we replaced 
$g_Z$ with $e_q$, set 
$F(0) \to 1$,  and 
multiplied by $2$ because each of the two photons can be produced in the fragmentation of the collinear $q\bar q$ pair.
For the perturbative short-distance amplitude of the  $H \to \gamma \gamma$ decay, first 
computed in Refs~\cite{Ellis:1975ap,Shifman:1979eb},
we  use Eq.~(\ref{eq3.13}), where 
we make a replacement  $g_Z \to \akh{e_q}$ 
for obvious reasons. 
   The result in Eq.~(\ref{eq4.1}) implies that the ratio 
of the non-perturbative amplitude to the perturbative one in the $H \to \gamma \gamma$ case  is nearly identical to Eq.~(\ref{eq:ratioZZ}).

The  physical picture 
of the non-perturbative corrections to the $H \to \gamma \gamma$ amplitude consistent with this result can be formulated as follows. 
 The long-distance  fragmentation of the collinear
 $q \bar q$ pair to 
 a photon is the main source  of the leading non-perturbative correction to the $H \to \gamma \gamma$ decay. However, only \emph{one} of the two photons in the decay is produced by this mechanism, whereas the second photon is produced at short distances.  Hence, the production of the second photon is  not subject to an 
 additional power  enhancement 
 by the ratio of the square of the short-distance scale to the hadronic scale, represented by the magnetic 
 susceptibility of the QCD vacuum. 
 
It is exactly this point that distinguishes our analysis from the enhancement mechanism discussed in 
 Ref.~\cite{Knecht:2025nyo}, since in that reference the  long-distance enhancement is postulated 
 for \emph{both} photons.   Although 
 we  believe that our analysis  of 
 the non-perturbative effects in the  $H \to \gamma Z$ amplitude  is better motivated than the discussion in Ref.~\cite{Knecht:2025nyo}, 
 there is no doubt that the result in Ref.~\cite{Knecht:2025nyo}
 can be used to provide an order-of-magnitude estimate of the non-perturbative effects in 
 the $H \to \gamma Z$ case  and, numerically, our results are similar. 
 However, for the $H \to \gamma \gamma$ decay, our result 
 is \emph{smaller} than the result in  Ref.~\cite{Knecht:2025nyo}
 by a factor $(\Lambda_{\rm QCD})^2/m_H^2 \sim 
 10^{-4}$ and,  therefore, it is  more in line with the  findings in Ref.~\cite{Haisch:2025bat}.

\section{Conclusions}

We have discussed the leading non-perturbative corrections to the $H \to  \gamma Z$ and $H \to \gamma \gamma$ decays.  These corrections  originate from 
 the light-quark loop contributions, 
 see Fig.~\ref{fig:diagL}.
 It is peculiar that, in contrast to  the regular perturbative light-quark short-distance 
 contributions 
 to these decays, which are suppressed by two powers of the light-quark mass, the non-perturbative effects are only suppressed by one power of $m_q$.  This was pointed 
 out earlier in Ref.~\cite{Knecht:2025nyo}, 
 and our analysis supports these findings. 
   
We have shown that one can use 
the well-established method of the operator product expansion on the light cone, to estimate the  non-perturbative corrections to the $H \to \gamma Z$ and $H \to \gamma \gamma$ decays.  The leading non-perturbative contributions  
are determined by the twist-two photon distribution amplitude, the quark condensate and the magnetic susceptibility of the 
QCD vacuum. Our analysis suggests that the \emph{leading} non-perturbative correction to the $H \to \gamma \gamma $ decay amplitude, originates from kinematic configurations where  one photon is produced by a long-distance fragmentation of the  $q \bar q $ pair, and the second one is produced at short distances. 
While
one \emph{can} identify  non-perturbative contributions  to $H \to \gamma \gamma$ decay where 
\emph{both} photons are produced at long distances, our analysis shows that they will be suppressed by  $ \Lambda^2_{\rm QCD}/m_H^2 \sim 10^{-4}$ relative  to the 
leading non-perturbative mechanism  established above. 

Numerically, the non-perturbative  effects are tiny.  They  modify  the 
$H \to \gamma Z$ and 
 $H \to \gamma \gamma$ 
decay rates  by about $10^{-5}$. Similar level of suppression 
was observed in Ref.~\cite{Haisch:2025bat} which utilized the dispersion 
relation  for the  $H \to \gamma \gamma$ form factor to estimate the non-perturbative 
corrections.   We therefore conclude that  non-perturbative  corrections to the $H \to \gamma Z$ and $H \to \gamma \gamma$
decays  are \emph{not an obstacle}  for the exploration  of these processes  at the  high-luminosity 
LHC and at future colliders with a percent-level 
precision. 

\section*{Acknowledgments}
The research of A.K. and K.M.  was supported  by  
the Deutsche Forschungsgemeinschaft (DFG, German Research Foundation) under grant no.\ 396021762 - TRR 257. 
The research of A.V. was supported in part by grant NSF PHY-2309135 to the Kavli Institute for Theoretical Physics (KITP). A.V. would like to acknowledge the long-term 
hospitality of the KITP.

\newpage

\bibliographystyle{JHEP}
\bibliography{newbib}

\end{document}